\documentclass[aps,prb,amsfonts,amssymb,longbibliography,twocolumn,amsmath,floatfix,showpacs]{revtex4-2}

\usepackage[dvips]{graphics}
\usepackage{graphicx} 

\usepackage[dvipsnames]{xcolor}
\usepackage{hyperref}%
\usepackage{xcolor}%
\definecolor{color}{HTML}{0000FF}
\hypersetup{linkcolor=color,urlcolor=color,citecolor=color,filecolor=color,colorlinks=true}

\usepackage{subfig}

\usepackage{amsbsy}

\usepackage{bm} 
\usepackage[T2A]{fontenc}
\usepackage[cp1251]{inputenc}
\usepackage[english]{babel}

\newcommand{\be}{\begin{equation}}
\newcommand{\ee}{\end{equation}}
\newcommand{\bel}{\begin{align}}
\newcommand{\eel}{\end{align}}
\def\bml#1\eml{\begin{multline}#1\end{multline}}
\def\bal#1\eal{\begin{align}#1\end{align}}
\def\bald#1\eald{\begin{aligned}#1\end{aligned}}

\newcommand{\beq}{\begin{equation}}
\newcommand{\eeq}{\end{equation}}
\newcommand{\bea}{\begin{eqnarray}}
\newcommand{\eea}{\end{eqnarray}}
\newcommand{\p}{\partial}

\DeclareMathOperator{\Img}{\mathrm{Im}}

\begin{document}
\title{Casimir-Lifshitz interaction between bodies integrated in a 
microelectromechanical/nanoelectromechanical quantum damped oscillator}

\author{Yu.\,S.~Barash}

\affiliation{Osipyan Institute of Solid State Physics RAS, 2 Academician Osipyan Street,
Chernogolovka, Moscow Region 142432, Russia}

\date{December 9}

\begin{abstract} 
A theory is proposed for the component of the Casimir-like force that arises between bodies embedded in a 
macroscopic quantum damped oscillator. When the oscillator's parameters depend on the distance between the 
bodies, the oscillator-induced Casimir-like force is generally determined by a broad spectral range 
extending to high frequencies, limited by the frequency dispersion of the damping function. 
Here it is shown that there exists a large class of systems in which the low-frequency range dominates the forces. 
This allows one to use the Ohmic approximation, which is crucial for extending the theory to a lumped
element description of fluctuation-induced forces in electrical circuits. Estimates of the circuit-induced 
Casimir-Lifshitz force suggest that under certain conditions it can be identified experimentally due to its 
dependence on various circuit elements. 
\end{abstract}

\maketitle                        

\section{Introduction}

Casimir forces originate from quantum fluctuations of the electromagnetic vacuum in the presence of 
boundaries or other inhomogeneities~\cite{Casimir1948,CasimirPolder1948}. Extending the theory to both 
quantum and 
thermal electromagnetic fluctuations in inhomogeneous equilibrium condensed matter systems, resulted in a 
general theory of Casimir and van der Waals (or Casimir-Lifshitz) forces between atoms, molecules, and 
macroscopic bodies~\cite{Lifshitz1955,DzyaloshinskiiPitaevskii1959,DzyaloshinskiiLifshitzPitaevskii1961} 
(see also reviews~\cite{BarashGinzburg1975,MahantyNinham1976,BarashGinzburg1989m,Parsegian2006,%
Bordagetal2009,Woodsetal2016}).

Advances in micro- and nanophysics technologies, particularly in mechanical transducers, atomic force 
microscopes, torsion pendulums, and oscillators, enabled detailed quantitative experimental
studies of Casimir-Lifshitz forces (see, for example,~\cite{Lamoreaux1997,MohideenRoy1998,DeccaetalKrause2003,%
Lamoreaux2005,Bordagetal2009,KlimchitskayaMohideenMostepanenko2009,RodriguezCapassoJohnson2011,Decca2015,%
RodriguezetalCapasso2015,Woodsetal2016,SomersetalMunday2018,GarrettetalMunday2019,Liuetal2019,%
LiuetalMohideen2019,GongetalMunday2021,Liuetal2021,BimonteetalDecca2021,SheldenSprengMunday2024} and 
references therein). These advances also drew significant attention to the influence of Casimir-Lifshitz 
forces on the performance of micro- and nanoelectromechanical devices, whose components 
often operate in close proximity~\cite{SerryWalliserMaclay1998,BuksRoukes2001,ChanetalCapasso2001,%
GuoZhao2004,LinZhao2005,Barcenasetal2005,EsquivelSirventetal2006,Batraetal2007,RamezaniAlastyAkbari2007,%
Ramezanietal2008,AndrewsetalTeufelLehnert2015,Sedighietal2015,Akhavanetal2019,StangeetalBishop2019,%
PateetalEdmund2020,JavoretalBishop2021,JavoretalBishop2022,BoucheetalBishop2024,Estesoetal2024,%
ElsakaetalBuhmannHillmer2024,Klimchitskayaetal2024,TajikPalasantzas2025}.

This paper proposes a theory of fluctuation-induced interactions between bodies embedded in 
quantum damped oscillators as encountered in micro- or nanoelectromechanical devices. The Casimir-like 
force considered here is driven by the oscillator through a parametric dependence of its frequency $\Omega$ 
and/or its damping function $\gamma(\omega)$ on the distance between the bodies. In electrical circuits, a 
similar Casimir-Lifshitz force component emerges from electromagnetic eigenmodes shaped by distance-dependent
lumped elements. This component can, under certain conditions, be detected experimentally, typically as a 
correction to the main Casimir-Lifshitz force. It can be distinguished by its dependence on various circuit 
elements.

The conventional microscopic approach to quantum effects in dissipative systems 
is based on the Hamiltonian of the entire conservative system, which includes the original system, its 
environment (thermal bath), and their interaction terms. The position and momentum correlation functions, 
the partition function, and some related thermodynamic quantities of the quantum damped oscillator have been 
studied in detail, mostly within the Zwanzig-Caldeira-Leggett model with 
bilinear coupling to the environment~\cite{Zwanzig1973,CaldeiraLeggett1983,GrabertWeissTalkner1984,%
GrabertSchrammIngold1988,HankeZwerger1995,Ingold2002,HaenggiIngold2005,HaenggiIngold2006,%
HaenggiIngoldTalkner2008,IngoldHaenggiTalkner2009,Weiss2012,SprengIngoldWeiss2013,TalknerHaenggi2020,%
Barash2021}. These quantities can, in particular, be represented as sums or products over the Matsubara 
frequencies $\omega_n=2\pi nT/\hbar$. 

The quantities in question can be divided into two groups, depending on the spectral range of fluctuations 
that dominate them. The first group, for example, includes the mean-square momentum fluctuation and the 
ground state energy. The corresponding sums diverge at high frequencies in the Ohmic approximation. 
A constant damping function over the entire spectral range implies the presence of a thermal bath with an 
infinite number of degrees of freedom, which links these divergences to those encountered in the 
quantum-field Casimir effect. Convergence then is ensured only by the frequency dispersion of the damping 
function $\gamma(i\omega_n)$, which is determined by the spectral density of the 
environment~\cite{CaldeiraLeggett1983,GrabertSchrammIngold1988,Weiss2012} decaying rapidly with
$\omega_n$ at high frequencies and defining a characteristic cutoff frequency range $\omega_n\alt\omega_c$ 
that governs the quantities in the first group. 

By contrast, the quantities of the second group are primarily determined by fluctuations in the spectral 
region $\omega\alt\tilde\Omega$, where $\tilde\Omega\sim\Omega$ for small or moderate $\gamma$. When 
$\omega_c\gg\tilde\Omega$, which is applicable to many problems and is assumed below, 
the condition $\omega\alt\tilde\Omega$ is favorable for applying the Ohmic approximation or a lumped 
element description. The mean squared position fluctuations of the damped oscillator and its specific heat are 
representative of the second group. Unlike the ground state energy, the specific heat 
belongs to the second group because it is determined by temperature derivatives of the 
thermal part of the energy~\cite{HaenggiIngoldTalkner2008,IngoldHaenggiTalkner2009,SprengIngoldWeiss2013}.

The oscillator-induced Casimir-like force provides an instructive example in this regard. Representing the 
distance derivative of the full free energy of the damped oscillator, the force, as shown 
below, belongs to the first group if the damping function depends on distance. In that case, the 
Casimir-like force includes a high-frequency contribution limited only by the frequency dependence of
$\gamma(i\omega_n)$, which must then be specified explicitly. The corresponding results will be obtained 
using the Drude model. 

The role of frequency dependent parameters is crucial for extending the theory of 
oscillator-induced Casimir-like forces to electrical circuits with a lumped element description. 
Divergences that generally arise when describing forces in lumped electrical circuits present a major 
theoretical challenge. Treating a quantum damped oscillator as a prototype for the electric-circuit 
eigenmodes, this paper identifies simple and general conditions under which finite (free of 
divergences) results for the corresponding contribution to the force can be obtained within the 
lumped-circuit description~\footnote{This paper considers only lumped circuits composed of lumped 
elements. In transmission-line circuits, effects of propagating waves along the circuit wires can 
substantially affect the results~\cite{Shahmoon2017,Leonhardt2018}.}. The results are derived from
the solution for a damped harmonic oscillator within the Zwanzig-Caldeira-Leggett model.

We show that the low-frequency range dominates the force only when the resonant frequency 
$\Omega$ depends on distance, whereas the damping function $\gamma$ does not. This enables the use of a 
lumped-element description of fluctuation-induced forces in electrical circuits. A similar description 
applies to cases where either $\gamma$, or both $\Omega$ and $\gamma$, depend on distance, but only $\Omega$ 
depends on an additional parameter $\varkappa$. In that case, the force difference 
$f_{\Omega_1} - f_{\Omega_2}$, with $\Omega_{1,2} = \Omega(\varkappa_{1,2})$, is governed by the 
low-frequency range. Explicit expressions for these forces are given below for two particular cases.

\section{Oscillator-induced Casimir-like forces}

Consider a quantum macroscopic damped oscillator, whose resonant frequency $\Omega$ and 
damping function $\gamma$ depend on the distance $d$ between the constituent interacting bodies. These 
quantities enter the quantum Langevin equation for the oscillator's coordinate 
$M\ddot{\hat{Q}}+M\int_{0}^{t}d\tau\gamma(t-\tau)\dot{\hat{Q}}(\tau)+M\Omega^2\hat{Q}=\hat{\xi}(t)$.  

From the free energy $F(d)=-T\log Z(d)$ of the damped oscillator in the Zwanzig-Caldeira-Leggett model, which
comes together with the partition function $Z(d)$~\cite{GrabertSchrammIngold1988,Ingold2002,Weiss2012,Barash2021}, 
it follows that the interaction force $f=-\tfrac{\p F}{\p d}$ can be written as~\footnote{See Supplemental 
Material at http://link.aps.org/ supplemental/10.1103/b6kg-nykc for details of the calculations of the oscillator-induced 
Casimir-like force under various conditions, as well as of the Casimir-Lifshitz force induced by the series 
and parallel RLC circuits. The Supplemental Material also contains 
Refs.~[\onlinecite{Bateman1,Abramowitz1972,Prudnikov2002}].}
\be
f=f_{\Omega}+f_{\gamma}=-T\sideset{}{'}\sum_{n=0}^{\infty}\frac{2\Omega(d)
\frac{\p\Omega(d)}{\p d}+\omega_n\frac{\p\gamma(i\omega_n,d)}{\p d}}{\omega_n^2+
\gamma(i\omega_n,d)\omega_n+\Omega^2(d)}.
\label{fOmegagamma}
\ee
Here $\gamma(\omega)=\int_0^\infty\gamma(t)e^{i\omega t}dt$, the summation is over Matsubara frequencies, 
and the prime at the summation sign indicates that the term with $n=0$ is taken with half weight. 

In \eqref{fOmegagamma}, each of the contributions, $f_{\Omega}$ or $f_{\gamma}$, describes repulsion, when 
$\Omega$ or $\gamma(i\omega_n)$ respectively decreases with distance, and describes attraction when it 
increases. 

The term $f_{\gamma}\propto\tfrac{\p\gamma(i\omega_n,d)}{\p d}$ belongs to the first group because it is 
sensitive to and limited by the frequency dependence of $\gamma(i\omega_n)$, and logarithmically diverges in 
the Ohmic approximation. By contrast, the term $f_{\Omega}\propto\tfrac{\p\Omega(d)}{\p d}$ is dominated by 
a comparatively low frequency range, $\omega_n\alt\Omega$ for small and moderate $\gamma$. For 
$\omega_c\gg\Omega$, the force component $f_{\Omega}$ can be treated in the Ohmic approximation and thus 
belongs to the second group. 

Focusing on oscillator- or circuit-induced forces shaped mostly by low-frequency fluctuations, we first 
assume that only $\Omega$ depends on $d$, while $\gamma(\omega)$ remains fixed (examples to follow). Then 
one obtains from \eqref{fOmegagamma} the following expression for the total fluctuation force 
$f(d)=f_{\Omega}(d)$ in the Ohmic approximation: 
\bml
f_{\Omega}(d)=-\left\{\dfrac{T}{\Omega}+\frac{i\hbar\Omega}{2\pi
\sqrt{\Omega^2-\frac{\gamma^2}{4}}}\left[\psi\biggl(1+i\frac{\hbar\omega_2}{2\pi T}\biggr) 
\right.\right. \\ \left.\left. -
\psi\left(1+i\frac{\hbar\omega_1}{2\pi T}\right)\right]\right\}\dfrac{\p\Omega}{\p d},
\label{drude72}
\eml
where the quantities $i\omega_{1,2}=\tfrac{\gamma}{2}\pm i\sqrt{\Omega^2-\tfrac{\gamma^2}{4}}$ enter the 
arguments of the digamma functions $\psi(x)$.

For weak dissipation $\gamma^2\ll4\Omega^2$, the force \eqref{drude72} is given by the derivative of
the free energy of the quantum harmonic oscillator $f=-\tfrac12\coth\tfrac{\hbar\Omega}{2T}
\tfrac{\p\Omega}{\p d}$. At low temperatures $T\ll \hbar\Omega$, this reduces to the zero-point energy
contribution $f_{\Omega}=-\tfrac{\hbar}{2}\tfrac{\p\Omega}{\p d}$. At high temperatures 
$\hbar\Omega\ll T$, the first term in \eqref{drude72} dominates the force. 

If $\gamma$ is not negligibly small, the dominant term that follows from \eqref{drude72} in the low 
temperature limit is
\be
f_{\Omega}=-\frac{\hbar\Omega\frac{d\Omega}{d d}}{\pi\sqrt{\gamma^2-4\Omega^2}}\ln\frac{\gamma+
\sqrt{\gamma^2-4\Omega^2}}{\gamma-\sqrt{\gamma^2-4\Omega^2}}.
\label{T0}
\ee

In the limit of a strong dissipation, $4\Omega^2\ll \gamma^2$, the frequency $i\omega_2\approx
\tfrac{\Omega^2}{\gamma}$ becomes anomalously small with increasing $\gamma$, while another one linearly 
increases $i\omega_1\approx\gamma$. The low temperature range is $T\ll\tfrac{\Omega^2}{\gamma}$ in this case, 
while the value of $\gamma$ is confined by the condition $\gamma\ll\omega_c$. 

\begin{figure*}%
\centering
\subfloat[][]{\includegraphics[height=0.200\textwidth]{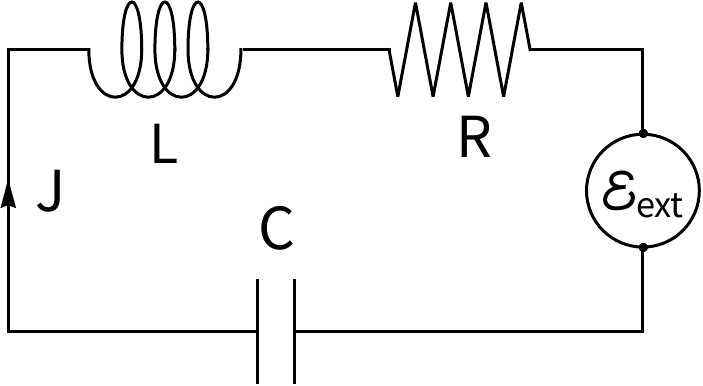}\label{fig1a}}%
\hspace{25mm}
\subfloat[][]{\raisebox{0.40cm}{\includegraphics[height=0.160\textwidth]{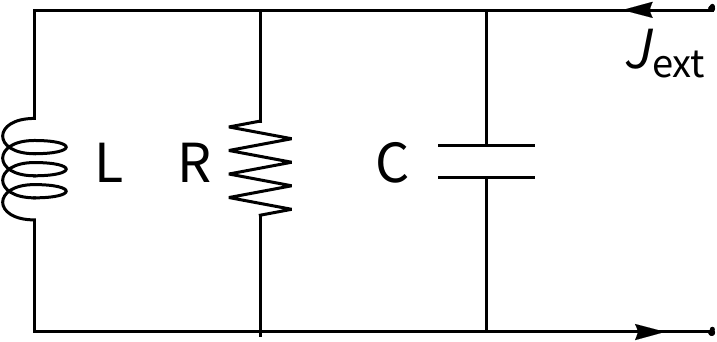}\label{fig1b}}}%
\caption{Simplest RLC circuits connected in series (a) and in parallel (b).}%
\label{figRLC}%
\end{figure*}

When $\gamma(\omega,d)$ depends on distance $d$, possibly along with $\Omega$, then an
additional term $f_\gamma\propto\tfrac{\p\gamma}{\p d}$ appears in the expression for the total fluctuation
force $f=f_\Omega+f_\gamma$, and a wide spectral range that includes the high frequences $\omega\alt\omega_c$
forms $f_\gamma$. However, if $\Omega$ depends on an additional parameter $\varkappa$, while 
$\gamma(i\omega_n,d)$ does not, then, as seen from from \eqref{fOmegagamma}, the difference 
$f_{\gamma}(d,\varkappa_1)-f_{\gamma}(d,\varkappa_2)$ belongs to the second group, whereas the individual 
quantities $f_{\gamma}(d,\varkappa_{1,2})$ get included into the first group.

Introducing the quantity $\tilde{f}(d,\varkappa)$, that satisfies the relation $\tilde{f}(d,
\varkappa_1)-\tilde{f}(d,\varkappa_2)=f(d,\varkappa_1)-f(d,\varkappa_2)$ and retains only those 
terms from $f(d,\varkappa)$, which generally are not canceled out in this difference, we obtain within the 
Ohmic approximation
\bml
\tilde{f}=-\dfrac{T}{\Omega}\dfrac{\p\Omega}{\p d}+\frac{\hbar}{4\pi}\dfrac{\p\gamma}{\p d}
\sum_{i=1,2}\psi\left(1+i\frac{\hbar\omega_i}{2\pi T}\right) + 
\\
\left[\psi\left(1+i\frac{\hbar\omega_1}{2\pi T}\right)-\psi\left(1+i\frac{\hbar\omega_2}{2\pi T}\right)
\right]\dfrac{i\hbar\left(\Omega\frac{\p\Omega}{\p d}-\frac{\gamma}4\frac{\p\gamma}{\p d}\right)}{2\pi
\sqrt{\smash[b]{\Omega^2-\frac{\gamma^2}{4}}}}.
\label{osc33nnn}
\eml

For describing the total oscillator-induced Casimir-like force in the presence of distance-dependent 
$\gamma$, the frequency dependence of $\gamma$ must be specified in \eqref{fOmegagamma}. Consider here the 
Drude-like model $\gamma(\omega)=\tfrac{\gamma_0\omega_D}{\omega_D-i\omega}$, which is commonly used in 
studying the damped oscillator properties~\cite{GrabertWeissTalkner1984,GrabertSchrammIngold1988,Ingold2002},
and assume that generally all three model parameters, $\Omega$, $\gamma_0$ and $\omega_D$, depend on $d$. 
A pronounced frequency dependence of $\gamma(i\omega,d)$ shows up in the model in the range $\omega\agt
\omega_D$, so that in \eqref{fOmegagamma} $\omega_D$ plays the role of an effective cutoff frequency 
$\omega_c$.

In the Drude model each term involving $\tfrac{\p\gamma_0}{\p d}$ or $\tfrac{\p\omega_D}{\p d}$ 
describes repulsion, when the parameter $\gamma_0$ or $\omega_D$ respectively decreases with $d$, and 
attraction when it increases.

When the frequency dispersion of $\gamma$ is disregarded, the damped oscillator has two complex 
eigenfrequencies $\omega_{1,2}$, as used in \eqref{drude72} and \eqref{osc33nnn}, whereas in the Drude model
it has three, $\omega_{1,2,3}$. If $\omega_D \gg \Omega, \gamma_0$, as is assumed here, these
frequencies can be readily obtained using an expansion in powers of the small parameters $\tfrac{\Omega}{\omega_D}$ 
and $\tfrac{\gamma_0}{\omega_D}$~\cite{GrabertWeissTalkner1984}. To the first order, $i\omega_{1,2}=
\tfrac{\gamma_0}{2}\pm i\sqrt{\Omega^2-\tfrac{\gamma_0^2}{4}}$ and $i\omega_3=\omega_D-\gamma_0\gg 
|\omega_{1,2}|$. In this case, the total fluctuation force is $f(d)=\tilde{f}(d)+\Delta f_\gamma(d)
$~\cite{Note2}. The first term is defined in \eqref{osc33nnn}, where $\gamma_0$ must be substituted for
$\gamma$, and the second term contains the third eigenfrequency as well as the parameter $\omega_D$: 
\bml
\Delta f_\gamma(d)=-\dfrac{\hbar}{2\pi}\psi\left(1+\frac{\hbar(\omega_D-\gamma_0)}{2\pi T}\right)
\dfrac{\p\gamma_0}{\p d}+ \\  \dfrac{\hbar}{2\pi}
\left[\psi\left(1+\frac{\hbar(\omega_D-\gamma_0)}{2\pi T}\right)-\psi\left(1+\frac{\hbar\omega_D}{2\pi T}
\right)\right]\dfrac{\p\omega_D}{\p d}.
\label{drude72a}
\eml

In the high temperature range $\omega_D\gg T\gg\Omega,\gamma_0$ one finds from \eqref{osc33nnn} and 
\eqref{drude72a}
\be
f=-\dfrac{T}{\Omega}\dfrac{\p\Omega}{\p d}
-\dfrac{\hbar}{2\pi}\dfrac{\p\gamma_0}{\p d}\ln\frac{\hbar\omega_D}{2\pi T}
-\dfrac{\hbar\gamma_0}{2\pi\omega_D}\dfrac{\p \omega_D}{\p d}.
\label{drude79}
\ee
Three contributions to the interaction force in \eqref{drude79} arise from the distance dependence of 
$\Omega$, $\gamma_0$ and $\omega_D$ and, in general, can have the opposite signs. The large argument under 
the logarithm sign in \eqref{drude79} reflects the logarithmic divergence of the sum in \eqref{fOmegagamma}, 
which appears when the frequency dispersion of $\gamma$ is neglected.

\section{Casimir-Lifshitz forces in circuits}

Assuming that dissipation in electrical circuits can be described within the 
Zwanzig-Caldeira-Leggett model~\cite{Devoret1997,BurkardKochDiVincenzo2004,VoolDevoret2017}, we apply the 
above results for the quantum damped oscillator to the damped eigenmodes of electrical circuits. To this end, 
we identify the resonant frequency and damping function for the particular circuit mode, and then 
substitute these in \eqref{fOmegagamma}--\eqref{osc33nnn}. In the general theory of van der Waals forces, 
the expansion over the damped eigenmodes is developed in detail~\cite{BarashGinzburg1989m}. Two particular 
RLC circuits, one in series and the other in parallel, are considered here as the simplest examples depicted
in Fig.~\ref{figRLC}.

In series RLC circuit, the current $J$ and the corresponding charge $Q$ ($J=\dot Q$) are 
common to all lumped elements. Therefore, it is convenient to take $Q$ as a generalised coordinate of the
system~\cite{LandauECM1984}. The corresponding canonical momentum that enters the Hamiltonian of the full
conservative system along with the coordinates and momenta of the environment, is the current-induced 
magnetic flux $\Phi=L\dot{Q}$. The effective equation of motion for the circuit, obtained after eliminating 
the environmental variables, is $L\ddot{Q}+R\dot{Q}+\tfrac{Q}{C}={\cal E}_{ext}(t)$, where 
${\cal E}_{ext}(t)$ is the external voltage. This equation describes the circuit dynamics and has the same 
form as that of a damped harmonic oscillator. In this analogy, $M\rightarrow L$, 
$\gamma\rightarrow\tfrac{R}{L}$ and $\Omega\rightarrow\tfrac{1}{\sqrt{LC}}$. Note that $\gamma$ here 
represents the frequency-independent Ohmic damping. 

If the capacitance $C$ depends on the distance $d$, while $R$ and $L$ do not, then only the oscillator 
frequency $\Omega$ depends on $d$, whereas $\gamma$ and $M$ do not. The contribution $f_{RLC}$ of 
electromagnetic fluctuations in the series circuit to the Casimir-Lifshitz force is then given 
by equation \eqref{drude72}, with parameters $\Omega_{LC}=\tfrac{1}{\sqrt{LC}}$ and $\gamma=\tfrac{R}{L}$. 

Alternatively, assume that the interacting bodies make a noticeable contribution to the circuit inductance, 
which therefore becomes dependent on the distance $d$ between them, while the capacitance is mainly 
determined by other circuit components. In this case, the capacitance can be varied at a fixed 
interaction, in particular, at fixed $L$. The finite force difference is then
$\tilde{f}_{RLC_1}-\tilde{f}_{RLC_2}$, where $\tilde{f}_{RLC}$ is defined in \eqref{osc33nnn} with 
$\Omega_{LC_{1,2}}=\tfrac{1}{\sqrt{LC_{1,2}}}$, $\gamma=\tfrac{R}{L}$ and 
$C_{1,2}=C(\varkappa_{1,2})$~\cite{Note2}. 
The parameter $\varkappa$ may, for example, correspond to two different distances between the 
capacitor plates (unrelated now to the interbody distance) or to two different values of the 
dielectric permittivity of the spacer.

In parallel RLC circuit, the applied voltage and the magnetic flux 
$\Phi=\int_{-\infty}^tV(t')dt'$ are common to all lumped elements. Therefore, it is convenient to choose 
$\Phi$ as the generalized coordinate of the system. The capacitor charge $Q_C=C\dot{\Phi}=CV$ serves as the 
canonical momentum in this case~\cite{Devoret1997,VoolDevoret2017,Rasmussenetal2021}. The effective equation 
of motion for the circuit, $C\ddot{\Phi}(t)+\tfrac1R\dot{\Phi}(t)+\tfrac{1}{L}\Phi(t)=J_{\text{ext}}(t)$, 
coincides with that of a damped harmonic oscillator, where $M\rightarrow C$, $\gamma\rightarrow\tfrac1{RC}$ 
and $\Omega\rightarrow\tfrac{1}{\sqrt{LC}}$. Here, as before, $\gamma$ is frequency-independent.

If the inductance $L$ depends on distance $d$, while $R$ and $C$ do not, then only the oscillator 
frequency $\Omega$ in the parallel RLC circuit depends on $d$, whereas $\gamma$ and $M$ remain unchanged. 
Therefore, the contribution $f_{RLC}$ of the circuit to the Casimir-Lifshitz force in the parallel 
configuration is determined by \eqref{drude72}, with parameters $\Omega_{LC}=\tfrac{1}{\sqrt{LC}}$ and 
$\gamma=\tfrac1{RC}$. 

Alternatively, assume that the interacting bodies make a noticeable contribution to the circuit capacitance, 
which thus becomes dependent on the distance $d$ between them, while the inductance is mainly determined by 
other circuit components. In this case, the $L$ can be varied independently at fixed $d$ and $C$. The finite 
force difference is then $\tilde{f}_{RL_1C}-\tilde{f}_{RL_2C}$, where $\tilde{f}_{RLC}$ is defined in 
\eqref{osc33nnn} with $\Omega_{L_{1,2}C}=\tfrac{1}{\sqrt{L_{1,2}C}}$, $\gamma=\tfrac1{RC}$, and 
$L_{1,2}=L(\varkappa_{1,2})$, corresponding, for example, to different spacing between the turns or to 
different numbers of the turns~\cite{Note2}.

We now estimate the Casimir-Lifshitz force component $f_{RLC}$ for a series RLC loop with distance-dependent 
capacitance and examine the ratio $r=\tfrac{f_{RLC}}{f_{Cas}}$, which compares 
this component to the main Casimir-Lifshitz force $f_{Cas}$ between the same bodies under similar 
conditions. 

Consider first a planar capacitor, for which the resonant frequency is 
$\Omega^{pc}_{LC}=\sqrt{\tfrac{d}{\varepsilon_0\varepsilon LS}}$, where $S$ and $d$ denote the contact area 
of the conducting plates and the distance between them, respectively; $\varepsilon_0$ is the vacuum 
permittivity. Focusing on the Casimir problem~\cite{Casimir1948}, we set $\varepsilon=1$ for the 
permittivity of the interlayer between the plates. 

The force at high temperatures described by the first term in \eqref{drude72}, $f_T=-\tfrac{T}{\Omega}
\tfrac{\p\Omega}{\p d}$, becomes $f^{pc}_{RLC,T}=-\tfrac{T}{2d}$ in this case. At low temperatures, 
one obtains from \eqref{T0} at weak ($R^2\ll\tfrac{4L d}{\varepsilon S}$) and strong 
($R^2\gg\tfrac{4L d}{\varepsilon S}$) dissipation, respectively:
\bal
&f^{pc}_{RLC,0}=-\frac{\hbar}{4\sqrt{\varepsilon_0 LS d}}+\frac{\hbar R}{4\pi L d}, 
\quad T\ll \frac{\hbar\sqrt{d}}{2\pi\sqrt{\varepsilon_0 SL}}, 
\label{rlc10}
\\
&f^{pc}_{RLC,0}=-\frac{\hbar}{2\pi \varepsilon_0 SR}\log\frac{\varepsilon_0 SR^2}{L d}, 
\quad T\ll\frac{\hbar d}{2\pi\varepsilon_0SR}.
\label{rlc1}
\eal

Given the applicability conditions of the lumped element approach, the strength of dissipation in 
\eqref{rlc1} is confined by the requirement $i\omega_1\approx\tfrac{R}{L}\ll\min(\omega_c, c/r_0)$, where 
$r_0$ is the typical size of a circuit element. 

The conventional Casimir force, $f_{Cas}=-\tfrac{\pi^2\hbar c S}{240 d^4}$, is the main 
fluctuation force component between metal plates in the strong retardation regime at low temperatures. At 
high temperatures, $T\gg \hbar c/d$, the force becomes $f_{Cas,T}=-\tfrac{\zeta(3)TS}{8\pi d^3}$. This 
expression follows from the Lifshitz theory with the Drude model~\footnote{As was noted long ago (see p.263 
in Ref.~[\onlinecite{Barash1988}]), this is half of what follows from the plasma model, since the 
latter fails to account for the complete suppression of the TM polarization contribution at high 
temperatures. This fact was later rediscovered and drew considerable attention due to a reported discrepancy
between precision measurements and predictions from Lifshitz theory~\cite{BostromSernelius2000,%
BordagGeyeretal2000,Bimonteetal2016,BimonteetalDecca2021,KlimchitskayaMostepanenko2023}}.

From these expressions we obtain the relative weights $r^{pc}_0$ and $r^{pc}_T$ of the circuit-induced 
Casimir-Lifshitz force between the plates of the planar capacitor in the dissipationless limit, 
$\hbar\gamma\ll \hbar\Omega, T$, at low and high temperatures:
\bal
&r_{0}^{pc}=\frac{60 d^{7/2}}{\pi^{2}c\sqrt{\varepsilon_0LS^{3}}}=
\frac{60}{\pi^2}\left(\frac{\Omega^{pc}_{LC}(d)}{c/d}\right)\frac{d^2}{S},
\label{r0} 
\\
&r_{\phantom{}_T}^{pc}=\frac{4\pi d^2}{\zeta(3)S}.
\label{rT}
\eal

As seen in \eqref{r0} and \eqref{rT}, the relative weight involves two small parameters at low temperature 
range, and only one at high. The quantities $r^{pc}_0$ and $r_{\phantom{}_T}^{pc}$, are both 
governed by the small geometric factor $\tfrac{d^2}{S}$. It reflects the~\mbox{different} dependence of the 
forces on the separation between the bodies and their characteristic sizes. Its smallness justifies 
neglecting edge effects at the boundaries of the plates. 

Another small parameter, $\tfrac{\Omega^{pc}_{LC}(d)}{c/d}$, that appears in $r_{0}^{pc}$ is the ratio 
of the oscillator frequency $\Omega^{pc}_{LC}$ and the characteristic frequency $c/d$ of quantum 
fluctuations forming the Casimir force. This ratio is typically small throughout the entire 
applicability domain of the Casimir result $f_{Cas}$. After switching over from low to high temperatures, 
both characteristic frequencies, $\Omega^{pc}_{LC}$ and $d/c$, present in \eqref{r0} are effectively 
replaced by $T$, which cancels out the frequencies in the relative weight, leading to \eqref{rT}, up to a 
numerical factor. 

As $\tfrac{d^2}{S}$ decreases from $0.04$ to $2.5\!\cdot\!10^{-3}$, the relative weight 
$r^{pc}_{\phantom{}_T}$ in \eqref{rT} drops from $0.42$ to $0.03$. The corresponding measurement accuracy 
requirements indicate that the force component under consideration can generally be detected at high 
temperatures.

For estimating the relative weights for a metal sphere and a flat metal plate, often used in experimental 
setups~\cite{Lamoreaux1997,MohideenRoy1998,Chanetal2001,ChanetalCapasso2001,DeccaetalKrause2003,%
DeccaetalKrause2003,Chanetal2008,deManetalIannuzzi2009,deManetalIannuzzi2010,SushkovetalLamoreaux2011,%
LaurentetalChevrier2012,Decca2015,Bimonteetal2016,Liuetal2019,StangeetalBishop2019,LiuetalMohideen2019,%
Liuetal2021,ElsakaetalBuhmannHillmer2024}, the capacitance could be computed numerically~\cite{Snow1954,%
Boyeretal1994,Crowley2008,Lekner2011}. However, for our purposes a simple interpolation 
$C^{sp-p}(d,R_{sp})=4\pi\varepsilon_0R_{sp}\left[1+\tfrac12\log\left(1+\tfrac{R_{sp}}{d}\right)\right]$
provides an adequate approximation for $d\alt R_{sp}$~\cite{Crowley2008}. Here, $R_{sp}$ is the sphere radius 
and $d$ is the minimum sphere-plate separation. In the dissipationless limit and at low temperatures, 
$T\ll \hbar \Omega^{sp-p}_{LC}$, the circuit-induced fluctuation force between the sphere and the plate, 
associated with the corresponding resonant frequency $\Omega^{sp-p}_{LC}=\tfrac1{\sqrt{LC^{sp-p}}}$, is
\be
f^{sp-p}_{LC}=-\frac{\hbar \Omega^{sp-p}_{LC}(d)}{8 d\bigl(1+\frac{d}{R_{sp}}\bigr)
\bigl[1+0.5\log\bigl(1+\frac{R_{sp}}{d}\bigr)\bigr]},
\label{estim1}
\ee 
while at high temperatures $T\gg \hbar\Omega^{sp-p}_{LC}$ one obtains
\be 
f^{sp-p}_{LC}=-\frac{T}{4d\bigl(1+\frac{d}{R_{sp}}\bigr)\bigl[1+0.5\log(1+\frac{R_{sp}}{d})\bigr]}.
\ee 

The main Casimir-Lifshitz attraction between the sphere and the plate, in the proximity force approximation, 
is given by $f^{sp-p}=-\pi^3 \hbar c R_{sp}/360 d^3$ at $T\ll \hbar c/d$, and $f^{sp-p}=-\zeta(3)T
R_{sp}/8d^2$ at $T\gg \hbar c/d$. Known refinements of these simple expressions~\cite{BimonteEmig2012,%
BimonteetalDecca2021} do not affect the subsequent conclusions.

Thus, in the dissipationless limit, the relative weights of the circuit-induced fluctuation force between 
the metallic sphere and plate at low ($T\ll \hbar c/d$) and high ($T\gg \hbar c/d$) temperatures are, 
respectively
\bal
&r^{sp-p}_0=\left(\frac{\Omega^{sp-p}_{LC}}{c/d}\right)\frac{1.45}{\bigl(\frac{R_{sp}}{d}+1\bigr)
\bigl[1+0.5\log\bigl(\frac{R_{sp}}{d}+1\bigr)\bigr]}, 
\label{rspp0} \\
&r^{sp-p}_T=\frac{1.66}{\bigl(\frac{R_{sp}}{d}+1\bigr)\bigl[1+0.5\log\bigl(\frac{R_{sp}}{d}+1\bigr)\bigr]}. 
\label{rsppT}
\eal

The factor $\Omega^{sp-p}_{LC}/(c/d)$ that enters \eqref{rspp0} is typically small within the applicability 
domain of the Casimir result. However, this factor is canceled out in \eqref{rsppT}. The geometric parameter 
$\tfrac{d}{R_{sp}}$ can vary here within comparatively wide limits. The relative weight $r^{sp-p}_T$ 
decreases from $0.5$ to $0.02$ as $\tfrac{d}{R_{sp}}$ decreases from $0.75$ to $0.035$. This shows a 
possibility to detect the circuit-induced force $f^{sp-p}_{LC}$ under optimal conditions.
The presence of multiple distance-dependent eigenmodes could enhance the effect.

\section{Discussion}

The lumped-element form of the equations of motion for electrical circuits is known to follow from a 
microscopic approach under certain conditions in a relatively low-frequency range. In this case, the circuit
parameters $C$, $L$ and $R$ can be regarded as phenomenological constants. To determine their specific
form and values in a given system constitutes an independent task.

As damped oscillators and dissipative circuits are open, nonconservative systems, a phenomenological 
description of their equilibrium energies and fluctuation-induced forces must, in general, be additionally 
validated by a microscopic theory. The Zwanzig-Caldeira-Leggett model provides the necessary framework for 
testing phenomenological schemes. It shows that averaging the conventional Hamiltonian of a free harmonic 
oscillator, $\tfrac{\langle P^2\rangle}{2M}+\tfrac M2\Omega^2\langle Q^2\rangle$, with the full equilibrium 
density matrix of the oscillator and its environment yields the correct expression for the energy of a damped
oscillator when the damping function is temperature-independent~\cite{HaenggiIngold2006,Barash2021}. In this 
case, the resulting expression for the free energy remains valid even in the presence of 
temperature-dependent damping $\gamma(\omega,T)$.

For distance-dependent parameters, both the averaged potential and kinetic energies become distance dependent
and jointly contribute to the interaction force. A similar statement is well known in the theory of van der 
Waals forces between atoms and molecules~\cite{FLondon1937}. The phenomenological approach of 
Ref.~[\onlinecite{BarashGinzburg1972}] to the fluctuation-induced energy of the series RLC circuit is based 
precisely on the expression $\tfrac12L\langle I^2\rangle+\tfrac{\langle Q_C^2\rangle}{2C}$, which includes 
both electric and magnetic contributions. The corresponding finite interaction force in the circuit was
obtained in Refs.~[\onlinecite{Barash1988}], \footnote{See the footnote on p.407 in 
Ref.~\cite{BarashGinzburg1989m}.} and shown to be consistent with the general theory of Casimir and van der 
Waals forces. This force coincides with \eqref{rlc10}, \eqref{rlc1} up to a minor typo~\footnote{In the limit
of large dissipation, described by equation (5.422) on page 248 in Ref.~[\onlinecite{Barash1988}], the 
resistance $R$ under the logarithm sign must be squared.} and therefore the theory developed in this 
manuscript supports the specific results obtained in 
Refs.~[\onlinecite{BarashGinzburg1972}],[\onlinecite{Barash1988}],[\onlinecite{Note4}]. 

An alternative phenomenological approach to Casimir-Lifshitz forces induced in electrical circuits 
was proposed in Ref.~[\onlinecite{Leonhardt2018}]. The principal difference from 
Refs.~[\onlinecite{BarashGinzburg1972}],[\onlinecite{Barash1988}],[\onlinecite{Note4}] is that 
Ref.~[\onlinecite{Leonhardt2018}] determines the forces solely from the potential term of the 
Hamiltonian, averaged over the state of the entire dissipative circuit, $\tfrac{\langle Q_C^2\rangle}{2C}$, 
neglecting all other Hamiltonian terms. This is inconsistent with the Zwanzig-Caldeira-Leggett model, 
insufficient for determining the forces, and leads to incorrect results which are in contradiction to those 
obtained here.

The discrepancy concerns both the explicit formulas for the forces in particular circuits, and 
the more general results and conclusions. One conclusion in Ref.~[\onlinecite{Leonhardt2018}] is that, in 
the dissipationless limit at zero temperature, the interaction potential differs from the zero-point energy.
However, as follows from the damped oscillator theory in this limit, the interaction potential coincides with 
the zero-point energy when all relevant Hamiltonian terms are included. Another conclusion in
Ref.~[\onlinecite{Leonhardt2018}] is that, within the lumped-element description, divergence of the 
interaction potential necessarily entails a divergence of the interaction force, even when it is only the 
resonant frequency that depends on distance. In particular, the expression for the interaction force in the 
parallel RLC circuit following from Ref.~[\onlinecite{Leonhardt2018}] diverges when $L$ is distance dependent 
at fixed $R$ and $C$. This contradicts our results \eqref{drude72} and \eqref{T0} when applied to that 
case (see also (S24)-(S27)~\cite{Note2}). In the series RLC circuit, the averaged potential 
$\tfrac{\langle Q_C^2\rangle}{2C}$ is finite and leads to a finite contribution to the
force~\cite{Leonhardt2018}, whereas the averaged magnetic energy $\tfrac12L\langle I^2\rangle$, omitted 
in Ref.~[\onlinecite{Leonhardt2018}], diverges within the lumped circuit description. In agreement 
with Refs.~[\onlinecite{Barash1988}],[\onlinecite{Note4}] and equations \eqref{rlc10}, \eqref{rlc1} (see 
also (S20)-(S23)~\cite{Note2}), the corresponding generally divergent conribution to the force 
becomes finite when $C$ is distance-dependent at fixed $R$ and $L$.

\section{Conclusions}

This paper develops a theory of Casimir-like forces induced by a quantum damped oscillator. Applied to bodies 
integrated in micro/nanoelectromechanical systems, this theory identifies the conditions, under which a 
lumped description of circuit-induced Casimir-Lifshitz forces is valid. Although the free energy diverges 
under these conditions, the circuit-induced component of the force is shown to be finite. The forces in 
series and parallel RLC circuits are described in detail. 

Two small parameters, one geometric and the other given by the ratio of characteristic frequencies,
determine the relative magnitude of circuit-induced fluctuation forces compared to the conventional 
Casimir-Lifshitz forces. Nevertheless, modern high-precision measurements make detailed studies of these 
forces feasible. Moreover, at high temperatures a single geometric parameter controls the relative magnitude 
of the effect, and under certain conditions the circuit-induced forces may constitute a substantial fraction 
of the total force.

The results obtained here can be straightforwardly extended to other problems involving damped oscillators or 
electrical circuits, in which derivatives of the free energy $F$ with respect to parameters other than the 
interbody distance are of interest. The only requirement is that the dependence of $F$ on the parameter be 
entirely determined by the dependence of $\Omega$ and $\gamma$ on it. For example, an interaction torque 
arises when $\Omega$ and/or $\gamma$ depend on a misorientation angle $\theta$ between the interacting 
bodies. This torque can be derived from \eqref{fOmegagamma}--\eqref{drude79} by replacing 
$\tfrac{\p\Omega}{\p d}$ and $\tfrac{\p\gamma}{\p d}$ with $\tfrac{\p\Omega}{\p\theta}$ and 
$\tfrac{\p\gamma}{\p\theta}$, respectively.

The work has been carried out within the state task of ISSP RAS.

\bibliography{mybib}

\renewcommand\thesection{\arabic{section}}
\setcounter{secnumdepth}{2}


\onecolumngrid
\clearpage
\begin{center}
{
{\large \textbf{Casimir-Lifshitz interaction between bodies integrated in a micro/nanoelectromechanical 
quantum damped oscillator}}
\\[10pt]
\textbf{Supplemental Material}
}
\thispagestyle{empty}

\vspace{0.15in}

Yu.\,S.~Barash

\vspace{0.075in}
\small\textit{Osipyan Institute of Solid State Physics RAS, 2 Academician Osipyan Street,
Chernogolovka, Moscow Region, 142432 Russia}

\end{center}

\vspace{3ex}

\numberwithin{equation}{section}

\twocolumngrid
\setcounter{subsection}{0}
\setcounter{equation}{0}
\renewcommand{\thesubsection}{S\arabic{subsection}}
\renewcommand{\theequation}{S\arabic{equation}}

\subsection{Casimir-like forces induced by the damped oscillator}
\label{sec: model}

The Zwanzig-Caldeira-Leggett model describes a harmonic oscillator coupled to a thermal bath, 
represented by a large (or infinite) set of other harmonic oscillators. In the absence of external forces, 
the Hamiltonian of the full system involving bilinear coupling with the environment takes the form
\bml
\hat{H}(d)=\dfrac{\hat{P}^2}{2M(d)}+\dfrac12 M(d)\Omega^2(d) \hat{Q}^2+ \\
\sum_{\alpha}\left[\dfrac{\hat{p}_\alpha^2}{2m_\alpha}+\dfrac12m_\alpha\omega_\alpha^2
\left(\hat{q}_\alpha-\dfrac{C_\alpha(d)}{m_\alpha\omega_\alpha^2}\hat{Q}\right)^2\right].
\label{osc8n}
\eml

Here, we additionally assume that the resonant frequency $\Omega(d)$ of the central oscillator, as well as, 
generally speaking, its mass $M(d)$ and its coupling constants with the surrounding oscillators 
$C_\alpha(d)$, depend on the distance $d$ between the bodies embedded in the scillator. This type of 
dependence can occur, in particular, in damped oscillators of micro-/nanoelectromechanical electrical 
circuits.

The interaction force is calculated in a statistically equilibrium state based on the Hamiltonian 
\eqref{osc8n} of the complete closed system, including the environment. The damping function enters the 
expression for the force only after statistical averaging. Since a small change in the distance between 
bodies, $\delta d$, is assumed to occur at nearly zero speed and at constant temperature, all changes in 
free energy, $\delta F$, should be attributed to the work of the interaction force 
along the path $\delta d$. After averaging, there is no dissipation in equilibrium.

Under these conditions, the interaction force is given by $f(d)=-\left\langle\frac{\p\hat{H}(d)}{\p d}
\right\rangle_T$, which, according to general results of statistical physics, is equivalent to $f=-\left(
\frac{\p F}{\p d}\right)_T$. This can also be verified by a direct calculation, which straightforwardly 
generalizes the calculation of the free energy of a damped oscillator in Ref.~[\onlinecite{Barash2021}] 
performed by integrating over the interaction constant.

When performing statistical averaging of the derivative of the Hamiltonian with respect to distance, the 
resulting spectral densities of the symmetrized correlation functions $\left(P^2\right)_\omega$, 
$\left(Q^2\right)_\omega$, and $\bigl(q_\alpha Q\bigr)_{\omega}$ can be described using the following 
fluctuation-dissipation relations
\be
\bigl(Q^2\bigr)_{\omega}=\hbar\coth\left(\dfrac12\beta\hbar\omega\right)\Img\Bigl[\chi(\omega)\Bigr],
\label{sri138}
\ee
\be
\bigl(q_\alpha Q\bigr)_{\omega}=\hbar\coth\left(\dfrac12\beta\hbar\omega\right)
C_\alpha\Img\Bigl[\chi(\omega)\chi_\alpha(\omega)\Bigr],
\label{sri140}
\ee
\be
\bigl(P^2\bigr)_{\omega}=\hbar\coth\left(\dfrac12\beta\hbar\omega\right)M^2\omega^2\Img\Bigl[\chi(\omega)\Bigr].
\label{sri141}
\ee 
Equations \eqref{sri138}-\eqref{sri141} involve the linear susceptibility of the damped oscillator
\be
\chi(\omega, d)=\dfrac{1}{M(d)
\left[\left(\Omega^2(d)-\omega^2\right)-i\omega\gamma(\omega,d)\right]},
\label{osc13n}
\ee
which depends on $d$ through the distance dependence of the damping function $\gamma(\omega,d)$, the resonant
frequency $\Omega(d)$, and the mass $M(d)$.

The damping function 
\be
\gamma(\omega,d)=-\,\dfrac{i}{\omega M(d)}\sum_\alpha C_\alpha^2(d)
 \left[\chi_\alpha(\omega)-\chi_\alpha(0)\right]
\label{osc14n}
\ee
depends on the interbody distance when the oscillator mass and/or its coupling constants
with the thermal bath modes exhibit such a dependence. 

At the same time, the susceptibility of the original individual reservoir mode (i.e., of the corresponding 
free oscillator) that enters \eqref{osc14n} and \eqref{sri140} does not depend on $d$:
\be
\chi_\alpha(\omega)=\dfrac{1}{m_\alpha}\dfrac{1}{\omega_\alpha^2-\omega^2-i\omega\varepsilon}, \quad 
\varepsilon\to+0\, .
\label{osc12n}
\ee

The calculation schematically outlined above leads to the equality $f=-\left(\frac{\p F}{\p d}\right)_T$, 
which involves the following expression for the free energy $F=-T\log Z$ of a damped oscillator
\be
F=T\ln\left[\dfrac{\hbar\Omega}{T}
\prod\limits_{n=1}^{\infty}\left(1+\dfrac{\Omega^2}{\omega_n^2}+
\dfrac{\gamma(i\omega_n)}{\omega_n}\right)\right].
\label{force5}
\ee
Free energy \eqref{force5} is known together with the corresponding partition 
function~\cite{GrabertSchrammIngold1988,Ingold2002,Weiss2012,Barash2021}. As a result, we arrive at the 
interaction force (1), given in the main text of this paper.

\subsection{Casimir-like forces in the Ohmic approximation}
\label{sec: Ohmicapp}

Suppose that the resonant frequency of the oscillator $\Omega$ varies with a certain parameter $\varkappa$, 
while the damping function $\gamma(\omega)$ does not depend on $\varkappa$. Let's consider the difference 
between free energies $F_{1,2}=F(\varkappa_{1,2})$, corresponding to two values $\varkappa_{1,2}$. It 
follows from \eqref{force5} 
\be
F_2-F_1=T\ln\left[\dfrac{\Omega_2}{\Omega_1}\prod\limits_{n=1}^{\infty}
\dfrac{\omega_n^2+\omega_n\gamma(i\omega_n)+\Omega_2^2}{\omega_n^2+\omega_n\gamma(i\omega_n)+\Omega_1^2}
\right],
\label{osc23nn}
\ee
where $\Omega_{1,2}=\Omega(\varkappa_{1,2})$.

Unlike \eqref{force5}, the expression on the right-hand side of \eqref{osc23nn} converges within the Ohmic 
approximation. This becomes obvious when the relation \eqref{osc23nn} is rewritten in the form
\be
F_2-F_1=T\sideset{}{'}\sum_{n=0}^{\infty}\ln\left(1+
\dfrac{\Omega_2^2-\Omega_1^2}{\omega_n^2+\omega_n\gamma(i\omega_n)+\Omega_1^2}\right).
\label{osc24nn}
\ee
Here, the prime at the summation sign indicates that the term with $n=0$ is taken with half-weight.

Within the Ohmic approximation adopted in this section, we use the roots $\omega_{1,2}$ of the equation 
\be
\omega^2+i\gamma\omega-\Omega^2=0,
\label{osc24nnp}
\ee
which represent the oscillator's complex eigenfrequencies in this approach.

One gets
\be
i\omega_{1,2}=\frac{\gamma}2\pm i\sqrt{\Omega^2-\frac{\gamma^2}{4}}. 
\label{osc25nn}
\ee

Given equation \eqref{osc25nn} and the Matsubara frequency $\omega_n=\tfrac{2\pi T}{\hbar}n$, 
the difference in free energies \eqref{osc23nn} can be written as
\be
F_2-F_1=T\ln\left[\dfrac{\Omega_2}{\Omega_1}\prod\limits_{n=1}^{\infty}
\dfrac{\left(n-\frac{\hbar\omega_{1}(\Omega_2)}{2\pi T}\right)
\left(n-\frac{\hbar\omega_{2}(\Omega_2)}{2\pi T}\right)}{
\left(n-\frac{\hbar\omega_{1}(\Omega_1)}{2\pi T}\right)
\left(n-\frac{\hbar\omega_{2}(\Omega_1)}{2\pi T}\right)}\right].
\label{osc28nn}
\ee

Equality $\omega_{1}(\Omega_2)+\omega_{2}(\Omega_2)=\omega_{1}(\Omega_1)+\omega_{2}(\Omega_1)$ ensures the 
convergence of the infinite product in \eqref{osc28nn}, allowing us to apply the summation formula (1.3.8) 
on p. 7 of Ref.~[\onlinecite{Bateman1}]. As a result, the quantity $F_2-F_1$ in \eqref{osc28nn} is expressed 
in terms of Gamma functions:
\be
F_2-F_1=T\ln\left[\dfrac{\Omega_2}{\Omega_1}\dfrac{\Gamma(1-\frac{\hbar\omega_{1}(\Omega_1)}{2\pi T})
\Gamma(1-\frac{\hbar\omega_{2}(\Omega_1)}{2\pi T})}{\Gamma(1-\frac{\hbar\omega_{1}(\Omega_2)}{2\pi T})
\Gamma(1-\frac{\hbar\omega_{2}(\Omega_2)}{2\pi T})}\right].
\label{osc29nn}
\ee

When moving from free energy to interaction force, two cases should be distinguished, both permitting the use 
of the Ohmic approximation. In the first case, the parameters $\varkappa$ and $d$ vary independently. In the 
second case, they coincide, $\varkappa\equiv d$.

Let only the frequency $\Omega$ depend on $\varkappa$ in the first case, while the damping function $\gamma$ 
depends on the distance $d$, possibly together with $\Omega$. Then the relatively low frequency range makes 
the dominating contribution to the difference of the interaction forces $f(\varkappa_2,d)-f(\varkappa_1,d)=-
\frac{\p}{\p d}(F_2-F_1)$. From \eqref{osc29nn} and \eqref{osc25nn}, it follows that this difference is 
described by expression (4) of the main text. The quantity $\tilde f(d,\kappa)$ in (4) satisfies the relation
$\tilde f(d,\varkappa_2)-\tilde f(d,\varkappa_1)=f(d,\varkappa_2)-f(d,\varkappa_1)$ and includes only those 
terms from $f(d,\kappa)$, which do not, in general, cancel out in the difference.

Considering the second case $d=\varkappa$, one assumes that only the resonant frequency $\Omega$ depends on 
$d$. Since not only the finite difference \eqref{osc29nn} between free energies, but also the derivative of 
the free energy with respect to $\varkappa$ can be described within the Ohmic approximation, one may set 
$d_2=d_1+\delta d$ in \eqref{osc28nn} and \eqref{osc29nn} and arrive at expression (2) of the main text for 
the total Casimir-like force, $f=-\left(\tfrac{\p F}{\p d}\right)_T$, induced by the oscillator.

In expanded form, this expression reads:
\begin{widetext}
\be
f=-\dfrac{T}{\Omega}\dfrac{\p\Omega}{\p d}- 
\left[\psi\left(1+\frac{\hbar\gamma}{4\pi T}-\frac{i\hbar}{2\pi T}\sqrt{\Omega^2-\frac{\gamma^2}{4}}\right)
-\psi\left(1+\frac{\hbar\gamma}{4\pi T}+\frac{i\hbar}{2\pi T}\sqrt{\Omega^2-\frac{\gamma^2}{4}}\right)
\right]\dfrac{i\hbar\Omega\frac{\p\Omega}{\p d}}{2\pi\sqrt{\Omega^2-\frac{\gamma^2}4}}.
\label{osc35nn}
\ee
\end{widetext}

In the limit of weak dissipation, $\gamma\ll\Omega, T$, we obtain from \eqref{osc35nn} keeping only the 
first-order terms in the expansion in powers of $\gamma$ and using the relation~\cite{Abramowitz1972}, 
$\Img\psi(1+iy)=-\tfrac{1}{2y}+\tfrac{\pi}{2}\coth(\pi y)$:  
\be
f=-\left[\frac{\hbar}2\coth\frac{\hbar\Omega}{2T}+\frac{\hbar^2\gamma}{4\pi^2 T}
\Img\left(\psi^{(1)}\biggl(1+\dfrac{i\hbar\Omega}{2\pi T}\biggr)\right)\right]
\frac{\p\Omega}{\p d},
\ee
where $\psi^{(1)}(z)=\tfrac{d\psi(z)}{dz}$.

In the high-temperature limit, we expand the $\psi$-functions in \eqref{osc35nn} in powers of small 
parameters, using the relations~\cite{Abramowitz1972} $\psi(1+z)=-\gamma_{eil}+\tfrac{\pi^2}{6}z+\ldots \,$ 
and $\psi^{(1)}(1+z)=\tfrac{\pi^2}{6}-2\zeta(3)z+\ldots$, which are valid at small z. Here 
$\gamma_{eil}=0.577\ldots$ is the Euler's constant. This leads to the equality
\be
f=-\left(\dfrac{T}{\Omega}+\dfrac{\hbar^2\Omega}{12T}\right)\dfrac{\p\Omega}{\p d}.
\label{drude76}
\ee

In the low temperature limit, it is convenient to transform \eqref{osc35nn} using the relation 
$\psi(1+z) = \tfrac{1}{z} + \psi(z)$ and to consider further the first terms of the asymptotic expansion 
$\psi(z)=\ln z-\tfrac1{2z}+\ldots$. This results in
\be
f=-\frac{\hbar\Omega\frac{d\Omega}{d d}}{2\pi\sqrt{\frac{\gamma^2}{4}-\Omega^2}}
\ln\frac{\frac{\gamma}{2}+\sqrt{\frac{\gamma^2}{4}-\Omega^2}}{
\frac{\gamma}{2}-\sqrt{\frac{\gamma^2}{4}-\Omega^2}}. 
\label{drude8487}
\ee
When $2\Omega\ge\gamma$, it is convenient to transform~\eqref{drude8487}~into
\bml
f=-\frac{\hbar\Omega\frac{d\Omega}{d d}}{2\sqrt{\Omega^2-\frac{\gamma^2}{4}}}
\left(1-\frac2{\pi}\arctan\frac{\frac{\gamma}{2\Omega}}{\sqrt{1-\frac{\gamma^2}{4\Omega^2}}}\right)
\label{drude84877}
\eml

\subsection{Casimir-Lifshitz forces induced in RLC circuits: one in series, the other in parallel}
\label{sec: circuit}

As shown in the main text, the interaction force, $f^s_{RLC}$, induced by the RLC circuit in series, can be 
obtained from \eqref{osc35nn}--\eqref{drude84877} by substituting $\Omega=\tfrac1{\sqrt{LC}}$, 
$\gamma=\tfrac{R}{L}$ and assuming that only the capacitance $C(d)$ depends on distance. As far as the RLC 
circuit in parallel is concerned, the interaction force $f^p_{RLC}$, induced by this circuit, is obtained 
from \eqref{osc35nn}--\eqref{drude84877} by substituting $\Omega=\tfrac1{\sqrt{LC}}$, 
$\gamma=\tfrac{1}{RC}$, when only the inductance $L(d)$ depends on distance.

It follows from \eqref{osc35nn} that the circuit-induced interaction force $f^s_{RLC}$ for the RLC circuit in 
series is
\begin{widetext}
\bml
f^s_{RLC}=\frac{T}{2 C}\frac{\p C}{\p d}+\frac{i\hbar\frac{\p C}{\p d}}{2\pi C^2 R 
\sqrt{\frac{4 L}{C R^2}-1}}
\left[\psi\left(1+\frac{\hbar R}{4\pi T L}\left(1-i\sqrt{\frac{4 L}{C R^2}-1}\right)\right) - \right. 
\\ \left. 
-\psi\left(1+\frac{\hbar R}{4\pi T L}\left(1+i\sqrt{\frac{4 L}{C R^2}-1}\right)\right)\right].
\label{circs1}
\eml

In the limit of weak dissipation  $R\ll \sqrt{\tfrac{L}{C}}, TL$ one gets
\be
f^s_{RLC}=\left[\frac{\hbar}{4\sqrt{LC^3}}\coth\frac{\hbar}{2T\sqrt{LC}}+\frac{\hbar^2R}{8\pi^2 
T(LC)^{\frac32}}\Img\left(\psi^{(1)}\biggl(1+\frac{i\hbar}{2\pi T\sqrt{LC}}\biggr)\right)\right]
\frac{\p C}{\p d}.
\label{circs2}
\ee
\end{widetext}
At high temperatures the force is
\be
f^s_{RLC}=\left(\frac{T}{2C}+\frac{\hbar^2}{24T LC^2}\right)\frac{\p C}{\p d}.
\label{drude76s}
\ee

In the low temperature limit we obtain
\be
f^s_{RLC}=\frac{\hbar\frac{d C}{d d}}{2\pi RC^2\sqrt{1-\frac{4L}{CR^2}}}
\ln\frac{1+\sqrt{1-\frac{4L}{CR^2}}}{1-\sqrt{1-\frac{4L}{CR^2}}}. 
\label{drude8487s}
\ee

Analogously, one gets for the RLC circuit in parallel
\begin{widetext}
\bml
f^p_{RLC}=\left\{\frac{T}{2 L}+\frac{i\hbar R}{2\pi L^2 \sqrt{\frac{4 C R^2}{L}-1}}
\left[\psi\left(1+\frac{\hbar}{4\pi TCR}\left(1-i\sqrt{\frac{4 C R^2}{L}-1}\right)\right) - \right. \right. 
\\ \left. \left.
-\psi\left(1+\frac{\hbar}{4\pi TCR}\left(1+i\sqrt{\frac{4 C R^2}{L}-1}\right)\right)\right]
\right\}\frac{\p L}{\p d},
\label{circs1p}
\eml

In the limit of weak dissipation  $R\gg \sqrt{\tfrac{L}{C}}, \tfrac1{TC}$ one gets
\be
f^p_{RLC}=\left[\frac{\hbar}{4L^{\frac32}C^{\frac12}}\coth\frac{\hbar}{2T\sqrt{LC}}+
\frac{\hbar^2}{8\pi^2 T(LC)^{\frac32}R}\Img\left(\psi^{(1)}\biggl(1+\frac{i\hbar}{2\pi T\sqrt{LC}}\biggr)
\right)\right]\frac{\p L}{\p d}.
\label{circs2p}
\ee
\end{widetext}

At high temperatures the force is
\be
f^p_{RLC}=\left(\frac{T}{2L}+\frac{\hbar^2}{24T L^2C}\right)\frac{\p L}{\p d}.
\label{drude76p}
\ee

In the low temperature limit we obtain
\be
f^p_{RLC}=\frac{\hbar R \frac{d L}{d d}}{2\pi L^2\sqrt{1-\frac{4CR^2}{L}}}
\ln\frac{1+\sqrt{1-\frac{4CR^2}{L}}}{1-\sqrt{1-\frac{4CR^2}{L}}}. 
\label{drude8487p}
\ee

\subsection{Casimir-like forces influenced by the frequency dispersion of the damping function}
\label{sec: Drude}

Within the Ohmic regime, the free energy expression \eqref{force5} diverges logarithmically at high
frequencies. The same divergence also occurs in (1) for the interaction force when the damping function 
depends on the interbody distance but not on frequency. In such cases, one must take into account the 
frequency dispersion of the damping function. As in other similar cases, solvable specific models of 
dispersion demonstrate possible behaviors of the convergent results~\cite{CaldeiraLeggett1983,%
GrabertWeissTalkner1984,GrabertSchrammIngold1988,Weiss2012}.

Here, we use the Drude model $\gamma(\omega)=\frac{\gamma_0\omega_D}{\omega_D-i\omega}$, assuming that the 
three model parameters $\Omega$, $\gamma_0$, and $\omega_D$ may depend on the distance $d$ between the 
bodies.
 
In this case, the free energy \eqref{force5} takes the form
\be
F=T\ln\left[\dfrac{\hbar\Omega}{T}
\prod_{n=1}^{\infty}\dfrac{(\omega_n+i\omega_1)(\omega_n+i\omega_2)(\omega_n+i\omega_3)}{
\omega_n^2(\omega_n+\omega_D)}\right],
\label{drude59}
\ee
and, as follows from expression (1) in the main text of the paper, the interaction force is 
$f=f_\Omega+f_{\gamma}=f_\Omega+f_{\gamma_0}+f_{\omega_D,1}+f_{\omega_D,2}$, where
\be
f_\Omega=-2T\Omega\frac{\p\Omega}{\p d}\sideset{}{'}\sum_{n=0}^{\infty}\frac{\omega_n +\omega_D}{
(\omega_n+i\omega_1)(\omega_n+i\omega_2)(\omega_n+i\omega_3)},
\label{drude59a}
\ee
\be
f_{\gamma_0}=-T\frac{\p\gamma_0}{\p d}\sum_{n=1}^{\infty}\frac{\omega_D\omega_n}{
(\omega_n+i\omega_1)(\omega_n+i\omega_2)(\omega_n+i\omega_3)},
\label{drude59b}
\ee
\be
f_{\omega_D,1}=-T\frac{\p\omega_D}{\p d}\sum_{n=1}^{\infty}\frac{\gamma_0\omega_n}{
(\omega_n+i\omega_1)(\omega_n+i\omega_2)(\omega_n+i\omega_3)},
\label{drude59c}
\ee
\bml
f_{\omega_D,2}=T\frac{\p\omega_D}{\p d}\times \\ \sum_{n=1}^{\infty}\frac{\omega_D\gamma_0\omega_n}{
(\omega_n+\omega_D)(\omega_n+i\omega_1)(\omega_n+i\omega_2)(\omega_n+i\omega_3)}.
\label{drude59d}
\eml

The complex eigenfrequencies of the oscillator $\omega_{1,2,3}$ appearing here satisfy the dispersion 
equation $\Omega^2-i\gamma(\omega)\omega-\omega^2=0$, which for the Drude model, takes the form
\be
\omega^3+i\omega_D\omega^2-(\Omega^2+\gamma_0\omega_D)\omega-i\Omega^2\omega_D=0.
\label{dispeq1}
\ee

The roots $\omega_{1,2,3}$ of this equation must satisfy the relations
\be
\left\{
\begin{aligned}
& \omega_1+\omega_2+\omega_3=-i\omega_D, \\
& \omega_1\omega_2+\omega_1\omega_3+\omega_2\omega_3=-(\Omega^2+\gamma_0\omega_D), \\
& \omega_1\omega_2\omega_3=i\Omega^2\omega_D.
\end{aligned}
\right.
\label{drude13}
\ee

In this model, it is generally assumed that the Drude frequency is much larger than the other 
characteristic frequencies of the problem $\omega_D\gg\Omega, \gamma_0$. Retaining the zeroth-order 
($\sim\omega_D$) and first-order ($\sim \Omega,\gamma_0$) terms yields an approximate 
solution~\cite{GrabertWeissTalkner1984}
\be\left\{
\begin{aligned}
&i\omega_1=\frac{\gamma_0}{2}+i\sqrt{\Omega^2-\dfrac{\gamma_0^2}{4}}, \\
&i\omega_2=\frac{\gamma_0}{2}-i\sqrt{\Omega^2-\dfrac{\gamma_0^2}{4}},\\
&i\omega_3=\omega_D-\gamma_0.
\end{aligned}
\right.
\label{dispeq24}
\ee

From the third equation in \eqref{drude13}, it follows that one of the eigenfrequencies, hereafter 
denoted $\omega_3$, is of zero-order; i.e., its absolute value is comparable to $\omega_D$. The values 
$|\omega_{1,2}|$ are the first order terms, $\alt\max\left\{\Omega,\gamma_0\right\}\ll \omega_D$.

Solution \eqref{dispeq24} satisfies the first relation in \eqref{drude13} exactly, and the second and third 
relations approximately. Identifying the second-order term on the right-hand side of the second 
equation in \eqref{drude13} requires knowledge of the second-order corrections to the second and third 
terms on its left-hand side: $\tilde{\omega}_1^{(2)}\tilde{\omega}_3^{(0)} +
\tilde{\omega}_2^{(2)}\tilde{\omega}_3^{(0)}$. Therefore, the approximate solution \eqref{dispeq24} reliably 
reproduces only the first-order term $-\gamma_0\omega_D$ in this relation. 

Regarding the third relation in \eqref{drude13}, solution \eqref{dispeq24} reliably reproduces the 
second-order term on the right-hand side, since the product of all three solutions on the left-hand side of 
\eqref{drude13} contains only zeroth- and first-order terms $\tilde{\omega}_1^{(1)}\tilde{\omega}_2^{(1)}
\tilde{\omega}_3^{(0)}$ and does not include second-order corrections $\tilde{\omega}_{1,2}^{(2)}$. The 
solution \eqref{dispeq24} satisfies the third relation in \eqref {drude13} only approximately, leading to 
extra third-order terms, because the product of the first-order terms $\tilde{\omega}_1^{(1)}
\tilde{\omega}_2^{(1)}\tilde{\omega}_3^{(1)}$ must be considered together with the products containing 
second-order corrections $\tilde{\omega}_1^{(2)}\tilde{\omega}_2^{(1)}\tilde{\omega}_3^{(0)}$ and 
$\tilde{\omega}_1^{(1)}\tilde{\omega}_2^{(2)}\tilde{\omega}_3^{(0)}$.

The validity of the first relation in \eqref{drude13} ensures the expression of the infinite product over 
Matsubara frequencies in \eqref{drude59} in terms of several Gamma functions (see, for 
example,~[\onlinecite{Bateman1}], p. 7, Sec. 1.3, (8)):
\be
\begin{aligned}
F&=-T\ln \left[\dfrac{T\Gamma\left(1+i\frac{\hbar\omega_1}{2\pi T}\right)
\Gamma\left(1+i\frac{\hbar\omega_2}{2\pi T}\right)\Gamma\left(1+i\frac{\hbar\omega_3}{2\pi T}\right)}{
\hbar\Omega\Gamma\left(1+\frac{\hbar\omega_D}{2\pi T}\right)}\right]\\ & 
=-T\ln \left[\hbar\Omega\dfrac{\Gamma\left(i\frac{\hbar\omega_1}{2\pi T}\right)
\Gamma\left(i\frac{\hbar\omega_2}{2\pi T}\right)\Gamma\left(i\frac{\hbar\omega_3}{2\pi T}\right)}{
4\pi^2 T\Gamma\left(\frac{\hbar\omega_D}{2\pi T}\right)}\right].
\end{aligned}
\label{drude62}
\ee

The second expression in \eqref{drude62} follows from the first one, due to the  equality $\Gamma(1+z)=
z\Gamma(z)$ and the third relation in \eqref{drude13}.

The expression for the oscillator-induced Casimir-like force 
$f=-\left(\frac{\p F}{\p d}\right)_{T=\text{const}}$ can be found most easily by directly differentiating 
expression \eqref{drude62}. Alternatively, one can sum over Matsubara frequencies in 
\eqref{drude59a}-\eqref{drude59d} using known relations (see, for example, Ref.~[\onlinecite{Prudnikov2002}],
p. 683, 5.1.24.6). Both methods lead to equivalent results within the accuracy of the approximation used. 
From the first expression in \eqref{drude62}, we find
\begin{widetext}

\be
\begin{aligned}
f=-\dfrac{T}{\Omega}\frac{\p\Omega}{\p d} 
-\frac{i\hbar}{2\pi}\left[\psi\biggl(1+i\frac{\hbar\omega_2}{2\pi T}\right.&\left.\!\!\biggr)- 
\psi\left(1+i\frac{\hbar\omega_1}{2\pi T}\right)\right]
\frac{\left(\Omega\dfrac{\p\Omega}{\p d}-\dfrac{\gamma_0}{4}\dfrac{\p\gamma_0}{\p d}\right)}{
\sqrt{\Omega^2-\frac{\gamma_0^2}{4}}}
+ \!\! \\ + 
\dfrac{\hbar}{4\pi}\left[\psi\left(1+i\frac{\hbar\omega_1}{2\pi T}\right)+\right.&\left.
\psi\left(1+i\frac{\hbar\omega_2}{2\pi T}\right)-
2\psi\left(1+\frac{\hbar(\omega_D-\gamma_0)}{2\pi T}\right)\right]\dfrac{\p\gamma_0}{\p d}+ 
\\
+\dfrac{\hbar}{2\pi}&\left[\psi\left(1+\frac{\hbar(\omega_D-\gamma_0)}{2\pi T}\right)
-\psi\left(1+\frac{\hbar\omega_D}{2\pi T}\right)\right]\dfrac{\p \omega_D}{\p d}.
\end{aligned}
\label{drude72s}
\ee

\end{widetext}
Formulas (4) and (5) in the main text describe the same result after replacing $\gamma$ with $\gamma_0$ in 
(3).

Using the expressions \eqref{drude62} and \eqref{drude72s}, the free energy and Casimir-like force can be 
easily described in the limiting cases of high and low temperatures, as well as for different possible 
relationships between temperature and Drude frequency. Here we present the corresponding results for the 
Casimir-like force. 

For a very high temperature $\Omega, \gamma_0\ll \omega_D\ll T$, when the parameter $\frac{\omega_D}{T}$ is 
small, one obtains from \eqref{drude72s} the following dominant terms in each of the three groups of terms, 
after expanding the digamma function near unity:
\be
f=-\dfrac{T}{\Omega}\dfrac{\p\Omega}{\p d}-\dfrac{\hbar^2\omega_D}{24T}\dfrac{\p\gamma}{\p d}
-\dfrac{\hbar^2\gamma}{24 T}\dfrac{\p \omega_D}{\p d}
\label{drude78}
\ee

If the temperature is high only in comparison to $\Omega$ and $\gamma$, while $\omega_D\gg 2\pi T\gg \Omega, 
\gamma$, then, in the presence of a large value of the argument $\frac{\omega_D}{2\pi T}$, the asymptotic 
expansion $\psi(1+z)=\ln z+\frac1{2z}+\ldots$ should be used. For the oscillator-induced Casimir-like force, 
this yields expression (6) in the main text.

At low temperatures, the arguments of all functions in \eqref{drude72s} take large values. 
Using the corresponding asymptotic expansions and keeping the dominant terms in each of the three groups of
terms containing derivatives $\frac{\p \Omega}{\p d}$, $\frac{\p \gamma_0}{\p d}$ and $\frac{\p \omega_D}{\p 
d}$, we arrive at the following expression:
\begin{widetext}
\be
f=-\frac{\hbar\Omega}{2\pi\sqrt{\frac{\gamma_0^2}{4}-\Omega^2}}
\ln\frac{\frac{\gamma_0}{2}+\sqrt{\frac{\gamma_0^2}{4}-\Omega^2}}{
\frac{\gamma_0}{2}-\sqrt{\frac{\gamma_0^2}{4}-\Omega^2}}\frac{\p\Omega}{\p d}
-\Biggl[\frac{\hbar}{2\pi}\ln\frac{\omega_D}{\Omega}
-\frac{\hbar\gamma_0}{8\pi\sqrt{\frac{\gamma_0^2}{4}-\Omega^2}}\ln\frac{\frac{\gamma_0}{2}
+\sqrt{\frac{\gamma_0^2}{4}-\Omega^2}}{\frac{\gamma_0}{2}-\sqrt{\frac{\gamma_0^2}{4}-\Omega^2}}
\Biggr]\frac{\p\gamma_0}{\p d}
-\,\frac{\hbar\gamma_0}{2\pi\omega_D}\frac{\p\omega_D}{\p d}.
\label{drude86} 
\ee

Under the condition $2\Omega>\gamma_0$, it is convenient to transform \eqref{drude86}
into the following form
\bml
f=-\frac{\hbar\Omega}{2\sqrt{\Omega^2-\frac{\gamma_0^2}{4}}}
\left(1-\frac2{\pi}\arctan\frac{\gamma_0}{2\sqrt{\Omega^2-\frac{\gamma_0^2}{4}}}\right)
\frac{\p\Omega}{\p d}-
\\
-\left[\frac{\hbar}{2\pi}\ln\frac{\omega_D}{\Omega}-\frac{\hbar\gamma_0}{8
\sqrt{\Omega^2-\frac{\gamma_0^2}{4}}}\left(1-\frac2{\pi}\arctan\frac{\gamma_0}{
2\sqrt{\Omega^2-\frac{\gamma_0^2}{4}}}\right)\right]\frac{\p\gamma_0}{\p d}
-\frac{\hbar\gamma_0}{2\pi\omega_D}\frac{\p\omega_D}{\p d}.
\label{drude86p} 
\eml

\end{widetext}

It can be verified that the factor preceding the term $\frac{\partial \gamma_0}{\partial d}$ 
in \eqref{drude86} is positive throughout the domain of applicability of the derived expression 
$\omega_D\gg \Omega, \gamma$. Therefore, in accordance with the result stated in the main text, each of the 
three contributions to the Casimir force in \eqref{drude86} represents attraction, when the corresponding
parameter $\Omega$, $\gamma_0$, or $\omega_D$ increases with $d$, and repulsion when it decreases with $d$.

\end{document}